\documentclass[aps,twocolumn,prl]{revtex4}

\input epsf
\def\plotone#1{\centering \leavevmode
\epsfxsize= 1.0\columnwidth \epsfbox{#1}}

\def\be{\begin{equation}}
\def\ee{\end{equation}}
\def\bea{\begin{eqnarray}}
\def\eea{\end{eqnarray}}

\def\cmm2{{\,\rm cm^{-2}}}
\def\cm2{{\,{\rm cm}^2}}
\def\cmm3{{\,{\rm cm}^{-3}}}
\def\gcmm3{{\,{\rm g\,cm^{-3}}}}

\def\fun#1#2{\lower3.6pt\vbox{\baselineskip0pt\lineskip.9pt
  \ialign{$\mathsurround=0pt#1\hfil##\hfil$\crcr#2\crcr\sim\crcr}}}

\def\vec{\bf}

\def\p3m{P$^3$M}

\def\fun#1#2{\lower3.6pt\vbox{\baselineskip0pt\lineskip.9pt
  \ialign{$\mathsurround=0pt#1\hfil##\hfil$\crcr#2\crcr\sim\crcr}}}

\def\micron{$\mu$m}

\begin{document}

\def\affilmrk#1{$^{#1}$}
\def\affilmk#1#2{$^{#1}$#2;}

\def\ucd{1}
\def\uc{2}

\bibliographystyle{prsty}
\title{Two Windows on Acceleration and Gravitation: Dark Energy or New Gravity?}
\author{Lloyd\ Knox\affilmrk{\ucd}\footnote{email:lknox@ucdavis.edu},
Yong-Seon\ Song\affilmrk{\uc}\footnote{email:ysong@cfcp.uchicago.edu},
J.Anthony\ Tyson\affilmrk{\ucd}\footnote{email:tyson@physics.ucdavis.edu}
}
\affiliation{
\parshape 1 -3cm 24cm
\affilmk{\ucd}{Department of Physics, One Shields Avenue
University of California, Davis, California 95616, USA}
\affilmk{\uc}{Department of Astronomy \& Astrophysics, 
University of Chicago, 5640 S. Ellis Avenue, Chicago, Illinois 60637, USA}
}
\date{\today}

\begin{abstract}
Small distortions in the observed shapes of distant galaxies, a cosmic
shear due to gravitational lensing, can be used to simultaneously
determine the distance-redshift relation, $r(z)$, and the density
contrast growth factor, $g(z)$.  Both of these functions are
sensitive probes of the acceleration.  
Their simultaneous determination
allows for a consistency test and provides sensitivity to
physics beyond the standard dark energy paradigm.
\end{abstract}
  \maketitle

{\parindent0pt\it Introduction.}
The observed acceleration of the cosmological expansion 
is driving a revolution in
fundamental physics.  This revolution could transform
our understanding of particles and fields (through the discovery of
a new ingredient, the ``dark energy'') or revise our deepest understanding
of space and time (by forcing fundamental changes to our theory
of gravity).  In this {\it Letter} we discuss how wide and deep 
tomographic cosmic shear surveys, through their sensitivity to both
geometry and the growth of density perturbations, can be used to 
distinguish between these two possibilities.  We also emphasize the
broader utility of having these two probes of, or windows on, acceleration
and gravitation.


Despite the variety of phenomena that can be explained with the 
cold dark matter cosmology, augmented with a dark energy component
\cite{riess98,perlmutter99,kaplinghat02b,white93,dodelson00,spergel03},
we still only know of dark energy through its gravitational influence.
And unlike dark matter, we have little hope of directly detecting the
dark energy via earth-bound laboratory experiments.  

Given that we only know of dark energy through its gravitational
effects, we must bear in mind the possibility that what we explain
with dark energy, may actually be due to corrections to Einstein gravity.
Note, as a historical precedent, that the anomalous perihelion
precession of Mercury detected in the 19th century was first explained
with unseen matter \citep{leverrier1860} before Einstein provided
the correct explanation.  

Assuming Einstein gravity, the growth of cold dark matter density
contrasts in the linear regime ($\delta({\vec x}, t) \equiv \delta
\rho({\vec x},t)/\bar \rho(t) << 1$) can be written as
$\delta({\vec x}, t)= g(t) \delta({\vec x}, t_i)$ where $t_i$ is some
early time and $g(t)$ is called the growth factor, usually
written as a function of redshift, $z$, instead.  The growth of
density contrasts results from a competition between the gravitational force pulling
matter toward overdensities and the expansion of the Universe
driving everything apart.  Thus $g(z)$ is sensitive to both the
gravitational force law and the history of the expansion rate.  With
the history of the expansion rate determined by $r(z)$, $g(z)$ can then
be used to test the gravitational force law on Mpc and larger scales
and thereby distinguish Einstein gravity from alternatives.

More generally, inconsistency of the standard dark energy paradigm
with the combination of $r(z)$ and $g(z)$ could arise for a variety of
reasons.  For example, the growth factor could also be altered by
non-gravitational interactions of the dark matter.  The cosmic-shear
inferred $r(z)$ and $g(z)$ may be internally consistent, but
inconsistent with $r(z)$ as inferred from supernovae due to
axion-dimming \cite{csaki02}.  It is thus imperative to probe geometry and
growth in as many ways as possible.  

{\parindent0pt\it Cosmic Shear Basics.}  Weak gravitational lensing
maps source galaxies to new positions on the sky, systematically
distorting their images. The resulting shear $\gamma$ of their images
is related to the projected foreground mass contrast inside an angular
radius $\theta$: \\ 
$\gamma_t = \bar \kappa(<\theta) - \kappa(\theta)$
where $\gamma_t$ is the tangential component of the shear, $\kappa$ 
is the mass surface density divided by  
$(c^2/4\pi G) (r_s / r_l r_{ls})$ and the $r_x$ 
are the angular diameter distances
of the source, lens, and lens-source \cite{miralda-escude91}.

By separating the galaxies into $n$ redshift bins, labeled
by $i$, we can create $n$ shear maps, $\gamma_i$.  The most
interesting statistical property of these maps, and the sole one we will
consider here, is the two-point function, $\langle \gamma_i \gamma_j \rangle$.
This two-point function is most easily expressed in the spherical
harmonic space in which we have 
$\langle \gamma_i^{lm} \gamma_j^{lm} \rangle = 
C_l^{ij} \delta_{ll'}\delta_{mm'}$. 

These $n(n+1)/2$ unique shear power spectra can be written as a 
projection of the matter power spectrum, $\Delta^2(k,z)$ along the 
line of sight:
\be
\label{eqn:shearint}
C_l^{ij} = \pi^2 l/2 \int dr r W(\bar r_i,r) W(\bar r_j,r)\Delta^2(k,z(r))
\ee
where 
$W(\bar r_i,r) \equiv \frac{\bar r_i -r}{\bar r_i r}$
for $r < r_i$ and zero otherwise \cite{bartelmann01,hu99b}.  Here 
$\bar r_i$ is the distance to galaxies in redshift bin $i$.  
The power spectrum is evaluated at the redshift
corresponding to distance $r$ on our past light cone and at wavenumber
given by $k = l/r$.  For simplicity we approximate the
distribution of galaxies in each redshift bin as a spike
of zero width at the center of the bin.
Below we assume $n=8$ redshift bins  
centered on $z_i = 0.2 + 0.4i$ where $i$ runs from 0 to 7.  
In the top panel of Fig.~1 we show the 
auto power spectrum for sources at $z = 1$.

{\parindent0pt\it Reconstructing $r(z)$ and $g(z)$: Qualitative.}
With cosmic shear maps from multiple source redshift bins, one can
simultaneously determine both $g(z)$ and the distance-redshift
relation, $r(z)$.  Counting the degrees of freedom one can see that
the multiple source redshift bins are essential: it would be
impossible to reconstruct two free functions from just a single shear
power spectrum.  Using multiple source redshift bins provides us with
the necessary further constraints.  

To gain further insight into the reconstruction, consider that the shear
power spectrum of a given source redshift bin is a sum of shear power from
lenses over a range of redshifts, as illustrated in the top panel
of Fig.~1.  Increasing $g(z)$ simply increases the
amplitude of the shear power contribution from structures at redshift $z$ by
$g^2(z)$.  Increasing $r(z)$
also changes the amplitude of the contribution from lenses at redshift $z$.
In addition it causes a shift of the power toward higher $l$ since the
three-dimensional structures, if further away, will project into smaller 
angular scales. For a single source redshift, the
changes in the shear power spectrum due to a change in $r(z)$ at just one
redshift, could also come from the appropriately chosen changes to $g(z)$
over a range in redshifts.  Thus it is impossible to simultaneously
reconstruct $g(z)$ and $r(z)$ from the shear power spectrum of a single
source redshift bin.  Including multiple source redshift bins breaks this
degeneracy.

If the matter power spectrum were a power law (and therefore all the
curves in Fig. 1 were power laws), then we would still have a
degeneracy between growth and distance even with multiple source
redshift bins.  The simultaneous determination is enabled by a bend in
the matter power spectrum, the exact location of which depends on the
matter density today, $\rho_m$ \footnote{Features introduced by non-linear
evolution play a subdominant role.}.  This feature, calibrated by CMB
determination of $\rho_m$, acts as a `standard ruler'
\cite{cooray01}.

\begin{figure}[htbp]
\label{fig:zbreakdown}
  \begin{center}
    \plotone{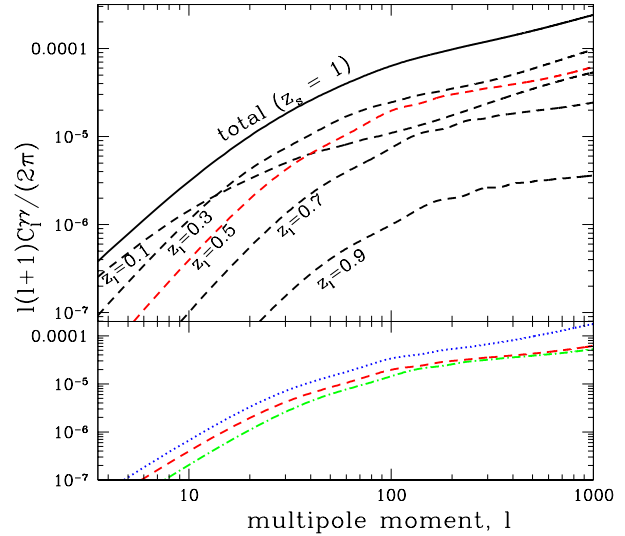}
    \caption{Dependence of the $z=1$ shear auto power
spectrum on $g(z)$ and $r(z)$.  There are $n-1$ more auto
power spectra and $n(n-1)/2$ cross power spectra not shown here.  
The solid line is the
shear power spectrum for sources at $z=1$.  The dashed lines show the
contributions to this shear power spectrum from lens slices of width
$\Delta z = 0.2$ centered at $z = 0.1, 0.3, 0.5, 0.7$ and 0.9.  Their
sum gives the solid line.  The lower panel shows the $z=0.5$ contribution
again (dashed line),
how it would look with an increase in $g(z=0.5)$ (dotted line)
and how it would look with an increase in $r(z=0.5)$ (dot-dashed line).
}
\end{center}
\end{figure}

{\parindent0pt\it Reconstructing $r(z)$ and $g(z)$: Quantitative.}
Our procedure is a modification of that in \cite{song05} where
simultaneous reconstruction of distances and growth factors from
cosmic shear data was first considered.

We parameterize $r(z)$ by 
its values specified at discrete redshift values $z_i = 0.4i$ for
$i=1$ to 8.  In addition we set $r(z_*) = r_s/\theta_s$ where
$z_*\simeq 1089$ is the redshift of last-scattering.  The sound
horizon at last-scattering, $r_s$, depends on $\rho_m$ and 
$\rho_b$.  These and the angular size of the sound horizon,
$\theta_s$, are constrained by CMB observations.
The values of $r$ at all other points of $z$ are found by linear
interpolation.  The $\Delta^2(k,z(r))$ factor in
Eq.~\ref{eqn:shearint} is evaluated by inverting $r(z)$
to get $z(r)$.

We assume that the primordial curvature power spectrum, 
with amplitude at wavenumber $k$
specified at horizon crossing (when $k/a = H$), is a quasi-power law with
logarithmically varying spectral index, $n_S(k) = n_S + \alpha_S \ln(k/k_f)$
where $k_f = 0.05$Mpc$^{-1}$.  This is the form of the expected power 
spectrum from
inflation.  The power spectrum at fixed time (or redshift) is related to
this primordial power spectrum by a scale-dependent transfer function, $T(k)$
and a growth factor, $g(z)$ so that
\be
\Delta^2_{\rm lin}(k,z) = \frac{2 \pi^2}{k^3} A_0 (k/k_f)^{n_S(k)-1} g^2(z) T^2(k) 
\ee
In general the time and scale-dependence do not factor as written here,
but we are interested in sufficiently small scales where the dark energy
perturbations can be ignored and in this case all modes grow at the same
rate.  The `lin' subscript on $\Delta^2$ here stands for `linear theory'.
We take non-linear evolution into account using 
the prescription of Peacock and Dodds~\cite{peacock96}.

We parameterize
$g^2(z)/a^2(z)$\footnote{We use the square to improve the Taylor expansion
approximation and divide by $a$
to improve the interpolation accuracy.} by its value at the eight discrete redshifts used for
parameterizing $r(z)$, plus its value at $z=0$.  We assume $g(z_*) =
a(z_*)$, as is the case in the matter-dominated era, and calculate 
$g^2(z)/a^2(z)$ at non-grid values of $z$ by linear 
interpolation.

The transfer function above depends on the matter content.  In
addition to $A_0$, $n_S$ and $\alpha_S$, the shear power spectra are
also therefore affected by $\rho_m$, $\rho_b$ and
the energy density of the cosmological neutrino background.  To
control these contributions we assume we have a measurement of the CMB
temperature and polarization power spectra as expected from Planck.
Since these CMB power spectra are also affected by the redshift of
reionization, $z_{ri}$, and the primordial fraction of baryonic mass
in $^4$He, $Y_p$, we include these parameters as well.  Our
parameter set is thus $\theta_s$, $A_0$, $n_S$,
$\alpha_S$, $\rho_m$, $\rho_b$, $z_{ri}$, $m_\nu$, $Y_P$),
eight $r(z)$ parameters and nine $g^2(z)/a^2(z)$ parameters.

To forecast errors, we Taylor expand to first order the dependence of
the shear power spectra on these parameters about our fiducial model.
The expected covariance matrix for the errors
in the estimated parameters can then be calculated via, e.g., Eq.~21
of \cite{song04}.  

{\parindent0pt\it Acceleration Without Dark Energy.}  As an example
of acceleration without dark energy we turn to
the `self-inflating' branch of the DGP model \cite{dvali00}. 
In this
model our 3+1-dimensional world (or `brane') is embedded in a 
4+1-dimensional space.  The Friedmann equation on the brane becomes
\be
H^2 -H/r_c = \frac{8\pi G}{3} \rho_m
\ee
and thus $H$ tends to a constant ($1/r_c$) as the Universe expands, just
as it would with the usual Friedmann equation in the presence of a cosmological
constant. 

The extra dimension also leads to a modification of the Poisson term
on scales between a smaller scale that is perhaps about 1 Mpc and 
a large scale, $r_c$.  Song \cite{song05} shows this to be 
\be
k^2/a^2 \Phi = \frac{3}{2}H^2
\left[\frac{1-H^{-1}/r_c}{1-H^{-1}/(2r_c)}\right] \delta
\ee
where $\delta \equiv \delta \rho_m/\bar \rho_m$ \footnote{See
\cite{nicolis04} for further discussion of fluctuations in the DGP
model.}.  Growth is suppressed in DGP gravity relative to dark matter
only Einstein gravity by the factor in square brackets.
However, dark energy also suppresses growth by $\bar \rho_m/\bar
\rho_{\rm tot}$, which, for Einstein gravity, replaces the factor in square
brackets.  In the following we set $r_c = 1.27/H_0$\cite{song05}.  
For non-linear
growth we apply the Peacock and Dodds prescription, as we do for
Einstein gravity, although consequences of the DGP model for
non-linear growth are at the moment unclear.

{\parindent0pt\it Quantitative Forecasts for a Fiducial Survey.}
We take the fiducial survey to be the ``2$\pi$'' deep wide survey of
20,000 square degrees in six wavelength bands from 0.4-1.1 \micron \
to be undertaken by the Large Synoptic Survey Telescope ($LSST$).
Several hundred sky-noise limited exposures in each optical band will
be obtained for each 10 square degree sky patch over a period of ten
years.  The shapes of galaxies out to a redshift of 3 (integrated
galaxy density of 50 per square arcminute) will be measured at a
precision far exceeding the 0.15 intrinsic random shear of an
individual source galaxy. These galaxy redshifts will be estimated
from fits of the 6-band fluxes to spectral templates for galaxies vs
type and redshift, a technique called photometric redshift estimation.

This survey will yield  the shear and redshift
of 3 billion source galaxies over a redshift range of 0.2 - 3.
Based on experiments with the active optics 8m Subaru telescope,
systematic shear error will be kept below 0.0001 on all angular
scales considered here.

Here we model the noise in the resulting shear maps as in
\cite{song04}.  Our analysis does not include effects of 
redshift errors. For systematic errors in distance
determinations to be less than 1\%, it is sufficient to require
$\Delta z/z < 0.01$ (since $r$ varies slower than linearly with $z$)
where $\Delta z$ is the error in the mean redshift of a given redshift
bin.  Simulations based on current
surveys indicate that this level of control is achievable at $z > 0.1$
\cite{connoly05}.  

Calculating $C_l^{ij}$ sufficiently accurately at
small scales is difficult \cite{zhan04,white04}.  We conservatively 
discarded data at $l > 1000$.  We have not modeled fluctuations in 
the dark energy, which can be important on large scales, and therefore
discard data with $l < 40$ as in \cite{song04}.  

{\parindent0pt\it Results.}
Our results are presented in Fig. 2.  To simulate reconstructed
$r(z)$ and $g(z)$ we add a realization of the errors, drawn from
a zero mean multivariate Gaussian with our forecasted covariance
matrix, to the fiducial values for $r(z)$ and $g(z)$.  The error bars 
are the square root of the diagonal
elements of this covariance matrix.  

\begin{figure}[htbp]
\label{fig:result}
  \begin{center}
    \plotone{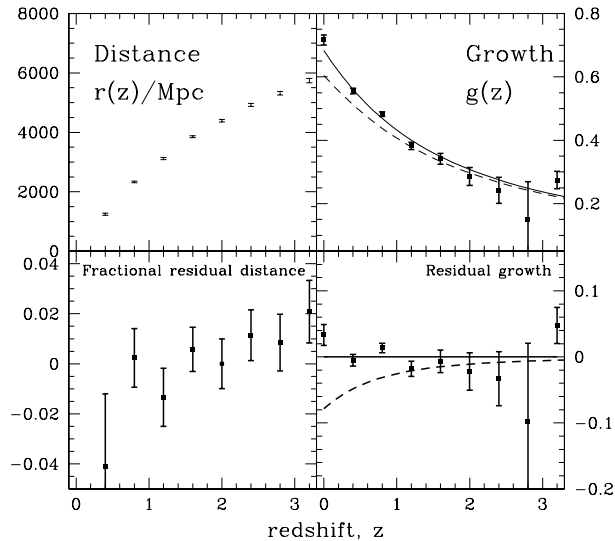}
    \caption{
Reconstructed distances (left panels), 
and growth factors (right panels).  The lower left panel shows the 
fractional residual distances, 
$[r(z)-r_{\rm fid}(z)]/r_{\rm fid}(z)$, where $r(z)$ are the reconstructed
distances and $r_{\rm fid}(z)$ are the distances in the fiducial DGP model.
The lower right panel shows the residual growth factor, $g(z)-g_{\rm fid}(z)$.
The curves in the right panels are $g_{\rm fid}(z)$ (solid) and 
$g(z)$ for the Einstein gravity model 
(dashed) with the same $H(z)$ and $\rho_m$ as the DGP model.  Although
these two models have the same $r(z)$ they are distinguishable by
their significantly different growth factors.}
\end{center}
\end{figure}

Distances are reconstructed with $\sim$ 2\% errors, even out to
$z=3.2$.  The distance errors are highly correlated; 
certain linear combinations will have even
smaller errors.  The growth factors have 3\% to 4\%  errors at $z \le 1.2$
and then grow steadily with $z$.
The tightness of the constraint at $z=3.2$ is an artifact
of our parameterization.  It is due to the $g^2(3.2)/a^2(3.2)$ parameter
influencing $g(z)$ all the way out to $z_*$ because of our interpolation
scheme.

One can distinguish dark energy from the DGP model even if the dark
energy density evolution is adjusted to match $r(z)$ for the DGP model.
As mentioned above, these two scenarios will
make different predictions for $g(z)$.  In the right panels the $g(z)$
curves for the fiducial DGP model and for the dark energy model with
identical $H(z)$ are shown.  Their difference in $\chi^2$ values is 
221, corresponding to almost $15\sigma$. 

As a test of our calculations,
we used our covariance matrix for $r(z)$ and $g(z)$ 
to calculate constraints on $w$, assuming a dark
energy model with constant $w \equiv P/\rho$.  
The results agree with a more direct calculation of the expected
error in $w$ that bypasses the $g$ and $r$ parameterization.  The
constraint on $w$ is almost entirely due to $r$ constraints rather
than $g$ constraints.  The roles of geometry and growth are
also discussed in \cite{zhang03,simpson04}.

We note that these measurements of distances into the matter-dominated
era, combined with Planck's CMB observations can be used to achieve
$\sigma((\Omega_{\rm tot}-1)h^2) \sim 10^{-2}$ \cite{knox05a}, greatly
improving the precision with which this robust prediction of inflation
can be tested.  Allowing for non-zero curvature will mean just one
more parameter to fit in our analysis and so will not qualitatively
degrade our eight distance determinations.

{\parindent0pt\it Discussion and Conclusions.} 
The statistical errors in our fiducial survey
are small enough to allow very precise reconstructions of distance
and growth as a function of redshift.  We have argued that
redshift errors will not qualitatively affect our results.

To investigate the impact of
shear calibration errors, we parameterized the observed $C_l^{ij}$ as
$f_i f_j$ times the true $C_l^{ij}$ and extended our parameter set to
include one gain parameter, $f_i$, for each source
redshift bin, $i$.  With a 1\% prior determination of all the calibration
parameters, we find that the distance errors increase by less than 25\%,
growth errors by less than 35\% and $\Delta \chi^2$
decreases from 221 to 137.
We expect to be able to determine the calibration
to even better than 1\% from
comparison of ground-based data with high-resolution space-based
images over a small fraction of the total survey area.  

The imaging data from which shear maps are derived can also be used to
infer galaxy power spectra.  The large-scale
feature from matter-radiation equality used here and the baryonic oscillations
at smaller scales can also be used to infer distances \cite{cooray01,seo03}.

The distance-redshift and growth-redshift relations provide two observational
windows on the physics of acceleration.  While we have illustrated the
utility of a second window with a specific example, the extra information 
may prove crucial to the unraveling of the mystery of 
acceleration in ways we have not yet imagined.  

\begin{acknowledgments}
We thank A. Albrecht, S. Aronson, G. Dvali, W. Hu, D. Huterer, 
N. Kaloper, R. Scoccimarro and C. Stubbs for  useful conversations.  
This work was supported at UCD by the National Science Foundation under 
Grants No. 0307961 and 0441072 and NASA under grant No. NAG5-11098 and at
UC by DoE No. DE-FG02-90ER-40560.
Data from the HST ACS and the Subaru telescope were used as input to
$LSST$ simulations.

\end{acknowledgments}

\bibliography{/work3/knox/bib/cmb3}

\end{document}